\documentclass[12pt,preprint]{aastex} 
\usepackage{graphicx} 
\def\etal{{\it et~al. }}

\newcommand{\ths}{\thinspace} 
 
\newcommand{\tle}{{\raise 0.3ex\hbox{$\sc {\ths \le \ths }$}}} 
\newcommand{\tge}{{\raise 0.3ex\hbox{$\sc {\ths \ge \ths }$}}} 
\newcommand{\tl}{{\raise 0.3ex\hbox{$\sc {\ths < \ths }$}}} 
\newcommand{\tg}{{\raise 0.3ex\hbox{$\sc {\ths > \ths }$}}} 
\newcommand{\ts}{{\raise 0.3ex\hbox{$\sc {\ths \sim \ths }$}}} 
\newcommand{\Msun}{\hbox{$\ths M_{\odot}$}} 
  
\shorttitle{Stability conditions of the HD$\thinspace$160691 
planetary system} 
\shortauthors{Bois \etal} 

\begin{document} 

\title{Conditions of Dynamical Stability for the 
HD$\thinspace$160691 Planetary System} 
\author{Eric Bois} 
\affil{Observatoire Aquitain des Sciences de l'Univers, Universit\'e 
Bordeaux 1, UMR CNRS/INSU 5804, B.P.89, F-33270 Floirac, France} 
\email{bois@obs.u-bordeaux1.fr}                   
\author{Ludmila Kiseleva-Eggleton} 
\affil{Dept. of Mathematics and Computer Science, 
St. Mary's College of California, Moraga, CA 94575, USA} 
\email{lkissele@stmarys-ca.edu}                   
\author{Nicolas Rambaux} 
\affil{OASU, Universit\'e Bordeaux 1, UMR CNRS/INSU 5804, B.P.89, 
F-33270 Floirac, France} 
\email{rambaux@obs.u-bordeaux1.fr} 
\and                   
\author{Elke Pilat-Lohinger} 
\affil{Institute of Astronomy, University of Vienna, 
T\"{u}rkenschanzstrasse 17, A-1180 Vienna, Austria} 
\email{lohinger@astro.univie.ac.at}                   

\begin{abstract} 
In our previous paper we showed that the currently determined
orbital parameters placed four recently announced planetary
systems HD$\thinspace$12661, HD$\thinspace$38529, 
HD$\thinspace$37124, and HD$\thinspace$160691 in very different
situations from the point of view of dynamical stability. In the
present paper, we deal with the last of these systems, whose
orbital parameters of the outer planet are yet uncertain. We
discover a stabilizing mechanism that could be the key to its
existence. The paper is devoted to the study of this mechanism
by a global dynamics analysis in the orbital parameter space
related to the HD$\thinspace$160691 system. We obtained our
results using a new technique called MEGNO and verified
them with the Fast Lyapunov Indicator technique (FLI).
In order to be dynamically stable, the HD$\thinspace$160691
planetary system has to satisfy the following conditions~: 
(1) a 2:1 mean motion resonance, 
(2) combined with an apsidal secular resonance, 
(3) in a configuration $P_{c}(ap) - S - P_{b}(ap)$ (which means
that the planets $c$ and $b$ may be considered as initially 
located at their apoastron around the central star $S$),
(4) and specific conditions on the respective sizes of the 
eccentricities. High eccentricity for the outer orbit 
($e_{c} > 0.52$) is the most probable necessary condition, 
while the eccentricity of the inner orbit $e_b$ becomes
relatively unimportant when $e_{c} > 0.7$. We also show that
there is an upper limit for planetary masses (in the interval
permitted by the undetermined line-of-sight inclination factor 
$\sin i_{l}$) due to the dynamical stability mechanism.  

More generally, in this original orbital topology, where the
resonance variables $\theta_{1}$ and $\theta_{3}$ librate about
$180^{\circ}$ while $\theta_{2}$ librates about $0^{\circ}$, 
the HD$\thinspace$160691 system and its mechanism have revealed
aspects of the 2:1 orbital resonances that have not been observed
nor analyzed before. The present topology with anti-aligned apsidal
lines combined with the 2:1 resonance is indeed more wide-ranging
than the particular case of the HD$\thinspace$160691 planetary
system. It is a new theoretical possibility suitable for a stable
regime despite relatively small semi-major axes with respect to
the important masses in interactions.
\end{abstract} 


\section{Introduction} 

In our previous paper (Kiseleva-Eggleton \etal 2002) we applied 
the new technique invented by Cincotta \& Sim\'o (2000) and 
called MEGNO (the acronym of {\it Mean Exponential Growth factor 
of Nearby Orbits}) to a wide neighbourhood of orbital parameters 
determined using standard two-body Keplerian fits for the 
recently discovered multi-planetary systems 
HD$\thinspace$12661, HD$\thinspace$38529, HD$\thinspace$37124, 
and HD$\thinspace$160691 in order to distinguish between regular 
and chaotic planetary configurations. We showed that the currently 
announced orbital parameters place these systems in very 
different situations from the point of view of dynamical 
stability. 
While HD$\thinspace$38529 and HD$\thinspace$37124 are located within 
large stability zones in the phase space around their determined 
orbits, the orbital parameters of the HD$\thinspace$12661 planets 
are located in a border region between stable and unstable dynamical 
regimes, so while its currently determined orbital parameters
produce stable orbits, a minor change within the margin of error
of just one parameter may result in a chaotic dynamical system. 
The orbits in HD$\thinspace$160691 (Jones \etal 2002) at first 
appeared highly unstable, but using MEGNO we were able to identify 
a few stability zones in a parameter space which included the 
parameters not determined from observations, such as the relative 
inclination $i_r$ between the two planetary orbits and the longitudes 
of the ascending nodes $\Omega$. All these stable configurations 
are associated with the 2:1 mean motion resonance. The present paper 
is wholly devoted to a detailed and complete dynamical analysis of 
the HD$\thinspace$160691 planetary system by taking into account the 
angular orbital parameters not constrained by observational data 
($i_r$, $\Omega$), as well as $\sin i_{l}$ (the line-of-sight
inclination factor) and the resulting different planetary masses. 
We have also explored the space of the mean anomalies $M$ of the
two planets taking initially into account the time of periastron
passage $\tau_{per}$ given by Jones \etal 2002 (see Table 1). 

In one of our earlier papers (Go{\'z}dziewski \etal 2002) we have 
clearly identified (using MEGNO) the exact location of the 2:1 mean 
motion resonance and its width for the Gliese 876 planetary system. 
A recent study by Hadjidemetriou (2002) of periodic orbits in this 
resonance predicts stable and unstable configurations of planetary 
systems depending on the hierarchy of planetary masses and 
eccentricities. The Gliese 876 system where $m_{b} < m_{c}$ and 
$e_{b} > e_{c}$ is, according to this study, a stable configuration. 
In contrast, the hierarchies of the HD$\thinspace$160691 system 
are inverse, i.e. $m_{b} > m_{c}$ and $e_{b} < e_{c}$.   
Hadjidemetriou (2002) found that a planetary system at the 2:1 
resonance where the inner planet is much more massive than the 
outer planet is unstable for all values of the eccentricities. 
However, in the HD$\thinspace$160691 system the two planetary 
masses are comparable and this leaves the question about its 
stability open. 

In this work we have explored the parameter space available 
for planets in HD$\thinspace$160691 in order to determine the 
stability conditions for this system. We have to notice that
the orbital parameters of both planets are rather speculative
due to the insufficient amount of observations, and even the
existence of the second planetary companion is not yet fully
confirmed (Jones \etal 2002, Butler \etal 2001). Nevertheless, 
the mechanism we present in this paper is probably the key to
the existence of planetary systems like HD$\thinspace$160691.
We have clearly identified the exact location of the 2:1 mean
motion resonance, its width, and the secular resonance in
apsidal longitudes preserving the stability related to the
mean motion resonance. 

\clearpage

\begin{deluxetable}{ccccccc} 
\tablewidth{0pt} 
\tablecaption{Orbital parameters of the HD$\thinspace$160691 
planetary system 
(Data from Jones \etal 2002, $M_*=1.08\Msun$) }

\tablehead{ 
\colhead{Planet}         & \colhead{$m_p\sin i_{l}$ ($m_J$)} & 
\colhead{$a$ (AU)}       & \colhead{$P$ (days)}              & 
\colhead{$e$} & 
\colhead{$\omega$ (deg)} & \colhead{$\tau_{per}$ (HJD)} } 

\startdata 
{\it b} & 1.7 $\pm$ 0.2 & 1.5 $\pm$ 0.1 & 
638 $\pm$ 10 & 0.31 $\pm$ 0.08 & 320 $\pm$ 30 & 50698 $\pm$ 30 \\ 
{\it c} & 1.0 & 2.3\tablenotemark{1} & 1300 & 
0.8 & 99 & 51613 \\ 

\enddata 

\tablenotetext{1} {In our numerical models we changed the 
value of $a_c$ from 2.3 AU to 2.381 AU (well within the 
error of its determination). The latter value gives the
exact location of the 2:1 resonance (by Kepler equation
resolution) and is located in the middle of the stability
valley on the $[a_{b}, a_{c}]$ parameter space (see Fig. 1a).} 
\end{deluxetable} 

\clearpage

\section{Method} 

A classical method that allows us to distinguish between 
regular and chaotic dynamical states is the method of Lyapunov 
Characteristic Numbers (LCN). The estimation of LCN usually 
requires computations over long evolutionary times, sometimes 
much longer than the lifetime of the system studied. Let us 
note that {\it chaotic} in the Poincar{\'e} sense means that 
the dynamical behavior is not quasi-periodic (the conventional 
definition useful for conservative dynamical systems), and
does not necessarily mean that the system will disintegrate
during any limited period of time. Let us state in addition
that we use the property of {\it stability} in the Poisson
sense~: stability is related to the preservation of a certain
neighbourhood relative to the initial position of the
trajectory. In conservative systems, quasi-periodic orbits
remain always confined within certain limits; in this sense
they are stable.

In the present work, we use two different methods in order 
to identify the dynamical state of the HD$\thinspace$160691 
system~: the {MEGNO} and {FLI} (the acronym of {\it Fast 
Lyapunov Indicator}) techniques. These two methods converge 
faster and are more sensitive than the LCN technique. 

FLI is the method introduced by Froeschl\'e \etal (1997), 
permitting us to distinguish qualitatively between regular 
and chaotic motion in a dynamical system (see for example 
Pilat-Lohinger \& Dvorak 2002). MEGNO is a new method 
developed by Cincotta \& Sim\'o (2000) that we have already 
successfully applied to the study of dynamical stability of 
extrasolar planetary systems (see e.g. Go{\'z}dziewski \etal 
2001, 2002, Kiseleva-Eggleton \etal 2002). This method 
provides relevant information about the global dynamics
and the fine structure of the phase space, and it yields 
simultaneously a good estimate of the LCN with a 
comparatively small computational effort 
(Cincotta \& Giordano 2000). The MEGNO is an alternative
technique that proves to be efficient for investigation
of both ordered and stochastic components of phase space
(Cincotta, Giordano \& Sim\'o 2002). It provides a clear
picture of resonance structures, location of stable and
unstable periodic orbits, as well as a measure of
hyperbolicity in chaotic domains (i.e. the rate of
divergence of unstable orbits) which coincides with
that given by the Lyapunov characteristic number. 

\placefigure{f1}

\clearpage

\begin{figure*} 
\epsscale{1.0}
%\plotone{f1.eps}
\caption{Stability maps in the $[a_b, a_c]$ parameter space 
in 2-D (a) and 3-D (b) for the HD$\thinspace$160691 planetary 
system (resolution of the grid is 30 x 30 points). 
In (a), filled and open circles indicate stable 
orbits ($<Y>=2 \pm 3\%$, and $<Y>=2 \pm 5\%$ respectively),
while small dots (not surrounded by circles) indicate highly
unstable orbits. $<Y>$ is the MEGNO indicator characteristic
value (Cincotta \& Sim\'o 2000).
(a) is a cross section of (b) in the plane $<Y>=2$ with 
projections of the different points according to their nature.
In (b), the peaks indicate the magnitude of instability. 
The stability strip in (a) corresponds to a stability
valley in 3-D (b) (however, its fine structure cannot 
be seen in this 3-D graph because of the scale factor in the
$z$ axis due to the high magnitude of the instability peaks).
In each Figure of the paper, the parameters used (besides 
those scanned) have the following values 
(initial conditions for numerical integrations)~:
$a_{b}=1.5$, $e_{b}=0.31$, $i_{b}=0^{\circ}$, 
$\Omega_{b}=0^{\circ}$, $\omega_{b}=320^{\circ}$, 
$M_{b}=156.3^{\circ}$; $a_{c}=2.381$, $e_{c}=0.80$, 
$i_{c}=1^{\circ}$, $\Omega_{c}=0^{\circ}$, 
$\omega_{c}=99^{\circ}$, $M_{c}=0^{\circ}$.}
\label{f1}
\end{figure*} 

\placefigure{f2}

\begin{figure}[ht]
\epsscale{0.5}
%\plotone{f2.eps} 
\caption{Stability map in the $[e_b, e_c]$ parameter space 
for the HD$\thinspace$160691 planetary system. The symbols 
and the grid resolution are the same as in all Figures 
(see Fig. 1a).} 
\label{f2}
\end{figure} 

\placefigure{f3}

\begin{figure*} 
\epsscale{1.0}
%\plotone{f3.eps}
\caption{Stability maps in the $[a_b, a_c]$ parameter space
plotted in 2-D (a) and 3-D (b) as in Fig. 1 and computed for
the HD$\thinspace$160691 system nominal elements (cf. Table 1
and Fig. 1) except for the mean anomaly of the {\it b} planet~: 
$M_{b}=0$. Although the 2:1 resonance in mean motions is 
formally preserved, the stability valley is lost.}
\label{f3}
\end{figure*} 

\clearpage
                   
\section{Stability conditions}                     		   
		   
In order to identify the different dynamical behaviors in the
parameter space, we use MEGNO which provides the exact location
of stable and unstable orbits as well as a measure of
hyperbolicity. Figures 1 and 2 show the dynamical state of
the HD$\thinspace$160691 system as a function of both orbital 
semi-major axes $a_b$ and $a_c$ (Fig. 1) and eccentricities 
{$e_b$} and {$e_c$} (Fig. 2). All other orbital parameters 
were taken from Table 1 with the addition of values of
$\Omega_{(b,c)}$ and $i_{(b,c)}$ undetermined from observations, 
and with $\tau_{per}$ replaced by the corresponding calculated 
values of mean anomalies $M=(2\pi/P)(t-\tau_{per})$ (with an 
initial time of reference $t=\tau_{per}^{c}$, one obtains 
$M_{b}=156.3^{\circ}$ and $M_{c}=0^{\circ}$). In order to
avoid a dynamical behavior bound to the plane, the two 
initial orbital inclinations $i_b$ and $i_c$ are taken
slightly different ($0^{\circ}$ and $1^{\circ}$ respectively); 
due to gravitational interactions of the 3-body problem, the
relative inclination $i_r$ is free to evolve in the 3-D space. 
In Fig. 2 as well as in other Figures in this paper $a_c$ was
taken to be 2.381 AU according to the Table footnote. In all
our Figures the intersection of horizontal and vertical lines 
indicates the `observational' initial parameters taken from 
Table~1. By using the MEGNO indicator characteristic value
$<Y>$ (Cincotta \& Sim\'o 2000), filled and open circles
in Fig. 1a (and all other Figures plotted in 2-D in this
paper) indicate stable orbits ($<Y>=2 \pm 3\%$, and
$<Y>=2 \pm 5\%$ respectively), while small dots (not
surrounded by circles) indicate highly unstable orbits
($<Y> \; \gg 2$).
One can see that the system is globally highly unstable,
except for a clearly marked stability region. 
Fig. 1a is a cross section of Fig. 1b (plotted in 3-D) in
the plane $<Y>=2$ with projections of the different points
according to their nature. In Fig. 1b, the peaks indicate
the magnitude of instability. 
The stability strip in Fig. 1a corresponds to a stability
valley in 3-D (Fig. 1b). However, its fine structure cannot 
be seen in this 3-D graph because of the scale factor in the
$z$ axis due to the high magnitude of the instability peaks. 

The stability strip is associated with the 2:1 mean motion
resonance. Generally, the presence of a resonance structures
the phase or parameter space in different regions, regular and
irregular (i.e. chaotic). In Fig. 1b, such different regions
appear clearly. The regions on both sides of the stability
valley present an irregular distribution of the values of 
$<Y>$ meaning very different magnitudes of instability. The
valley on the other hand presents a regular picture at the
used scale, which is confirmed by the cross section (a) of
(b). The strip indeed obtained in Fig. 1a, composed of
filled and open circles, is sufficiently homogeneous and
dense to be a stability zone. It is rather wide ($\sim 0.1$
AU in both semi-major axes), and besides one order of
magnitude larger than the one obtained for the Gliese 876
planetary system (also in 2:1 mean motion resonance, see
Go\'{z}dziewski, Bois, \& Maciejewski, 2002). 
However, for HD$\thinspace$160691 this stability valley 
may completely disappear for a small change in angular 
elements. The stability zone is indeed only permitted for
a parti\-cu\-lar geometrical configuration of the orbits 
(determined by the combination of the elements $\Omega$,
$\omega$, and $i_r$) combined with particular relative
positions of the two planets on their orbits determined
by the $M$ elements. For example, for a shift only on the
$M_{b}$ element without losing the 2:1 mean motion
resonance, close approaches at the periastron occur in
the end, and consequently the dynamical behavior becomes
totally and highly unstable as it is shown in Figure 3
(computed with $M_{b}=0^{\circ}$ instead of $156.3^{\circ}$).
The same situation appears with a comparable shift on the 
$\tilde{\omega}$ elements. For instance, with 
$\tilde{\omega}_{b}=0$, 100, or $200^{\circ}$ instead of 
$320^{\circ}$, the stability valley is lost. 
The whole stabilizing mechanism depends on particular
combinations of the elements $\Omega$, $\omega$, $i_r$, 
and $M$. We will discuss these stability conditions in
detail below. 

Figure 2 shows two stability regions in the $[e_b, e_c]$ 
parameter space~: (i) a small one with few points where both
eccentricities are small, and (ii) a much larger region for
high values of the outer orbit ($e_{c}>0.5$). Let us notice
that in the latter, the eccentricity of the inner orbit
$e_{b}$ is relatively unimportant when $e_{c}>0.7$. Let us
emphasize that the large region with a sizeable number of
stable points decreases significantly when $a_c$ is taken
to be 2.3 instead of 2.381 AU (see Fig. 3d in our previous
paper Kiseleva-Eggleton \etal 2002). In the present Figure 2,
due to the resolution of the grid, the dynamical behavior of 
the`observational' point in eccentricities (Table 1) does not
appear. However, with a higher resolution (i.e. more integration
points), computations confirm the existence of a stable point 
at the intersection of the horizontal and vertical lines.
In addition, few unstable points are also very close. 
Let us recall that some zones (notably on the sides of stability
structures) may present a sensitiveness to initial conditions
for stable versus unstable points in a very close vicinity
between them. This situation reminds us of the dynamical
state of the HD$\thinspace$12661 planetary system 
(Kiseleva-Eggleton \etal 2002). 
  
\placefigure{f4}

\clearpage

\begin{figure} 
\epsscale{1.0}
%\plotone{f4.eps} 
\caption{Stability maps in the parameter spaces 
$[\omega_{b}, i_{b}]$ (a), $[\Omega_{b}, i_{b}]$ (b), 
$[\omega_{c}, i_{c}]$ (c), and $[\Omega_{b}, i_{b}]$ (d) 
for the HD$\thinspace$160691 planetary system. The symbols 
and the grid resolution are the same as in all Figures 
(see Fig. 1a).} 
\label{f4}
\end{figure} 

\clearpage
  
Figure 4 shows how the dynamical state of the 
HD$\thinspace$160691 system depends on the relative inclination 
$i_{r}$ between the planetary orbits and on the angular elements 
$\omega$ and $\Omega$ for both planets $b$ and $c$. In the
general case of a 2-planet system, the relative inclination $i_{r}$ 
is related to the longitudes of ascending nodes $\Omega_{b}$ and 
$\Omega_{c}$ as follows~:
$$\cos i_{r}=\cos i_{b} \cos i_{c} + \sin i_{b} \sin i_{c} 
\cos(\Omega_{c}-\Omega_{b})$$
However, in our computations, for a variation of $i_{b}$ or $i_{c}$, 
the inclination of the other orbit is taken equal to zero. As a 
consequence, $i_{r}=i_{b}$ or $i_{c}$, and then the inclinations on
the vertical axes of the Figure 4 may be simply read as a scanning
of the relative inclination $i_{r}$ between the orbits. It is
important to remark that the distributions of structures are
relatively similar in the pairs 
$[\omega_{b}, i_{b}]$ and $[\omega_{c}, i_{c}]$, as well as in 
$[\Omega_{b}, i_{b}]$ and $[\Omega_{c}, i_{c}]$, 
with stable orbits appearing in strips of a comparable 
width $\sim 100^{\circ}$ around the nominal values of 
$\omega_b$ and $\omega_c$ (Table 1) for all values of $i_{r}$. 
There are also additional stability zones, but very narrow,
for small values of $i_r < 20^{\circ}$ in both maps related
to $\omega$, as well as in the two others related to $\Omega$.
The nominal values for $\omega_{b}$ and $\omega_{c}$, namely
320 and 99 degrees respectively, are located well inside the
stable strips. 

\placefigure{f5}

\clearpage

\begin{figure} 
\epsscale{0.50}
%\plotone{f5.eps} 
\caption{Stability map in the 
$[\tilde{\omega}_{b}, \tilde{\omega}_{c}]$ 
parameter space for the HD$\thinspace$160691 planetary system 
($\tilde{\omega}=\Omega+\omega$). The symbols and the grid 
resolution are the same as in all Figures (see Fig. 1a).} 
\label{f5}
\end{figure}

\placefigure{f6}

\begin{figure}
\epsscale{0.5}
%\plotone{f6.eps} 
\caption{Stability map in the $[M_b, M_c]$ parameter space 
for the HD$\thinspace$160691 planetary system. The symbols 
and the grid resolution are the same as in all Figures 
(see Fig. 1a).} 
\label{f6}
\end{figure} 

\clearpage

Figures 5 and 6 present the stability maps of the 
$[\tilde{\omega}_{b}, \tilde{\omega}_{c}]$ 
(where $\tilde{\omega}=\Omega+\omega$) and $[M_{b}, M_{c}]$
parameter spaces, respectively. Fig. 5 shows a linear
relationship between the longitudes of periastron 
$\tilde{\omega}_{b}$ and $\tilde{\omega}_{c}$ for stable
configurations. This means that $\tilde{\omega}_{b}$ 
and $\tilde{\omega}_{c}$ precess at the same rate. 
Fig. 6 shows stable strips in the mean anomaly $[M_b, M_c]$
plane that permit us to identify adequate relative
orbital positions for stable configurations at the 2:1
resonance in mean motions. Figures 5 and 6 also show that
the HD$\thinspace$160691 planetary system is dynamically
stable only when the corresponding orbital parameters are
simultaneously represented by stable points inside the
strips in both $[\tilde{\omega}_{b}, \tilde{\omega}_{c}]$ 
and $[M_{b}, M_{c}]$ planes. 
If only one of the two conditions is realized, the map 
$[a_{b}, a_{c}]$ exhibits a totally unstable dynamical 
state without any particular stability structure 
(see Fig. 3). We have also computed the 
$[\tilde{\omega}_{b}, \tilde{\omega}_{c}]$ stability maps
for different relative inclinations (from 0 to $90\deg$, 
not shown in this paper), and consequently we may point 
out that the whole mechanism of stabilization can be 
successfully applied even to high relative inclinations 
if $\Omega, \omega, M$ are determined in an appropriate way~: 
$\tilde{\omega}_{b}$ and $\tilde{\omega}_{c}$ on average
precess at the same rate while the planets {\it b} and
{\it c}, in 2:1 mean motion resonance, are located on
their orbits in such a way that there are no close
approaches at their periastron. The favourable respective
positions are defined by the stable strips in the mean
anomaly $[M_b, M_c]$ plane (cf. Fig. 6). 

The stability requires a particular geometrical configuration
of the orbits (defined by elements $\Omega$, $\omega$, and
$i_r$), added to particular positions of the two planets on
their orbits defined by the $M$ element. In addition to the
2:1 mean motion resonance, the planets {\it b} and {\it c}
have to be initially located in favourable respective
positions, for example at their apoapses,
in order to avoid close approaches at their periapses.
Following analogous properties analyzed by Lee \& Peale
(2002) for the GJ$\thinspace$876 planetary system, we have
found in the HD$\thinspace$160691 system the simultaneous
librations of the two mean motion resonance variables
$\theta_{1}=\lambda_{b}-2\lambda_{c}+\tilde{\omega}_{b}$ and
$\theta_{2}=\lambda_{b}-2\lambda_{c}+\tilde{\omega}_{c}$ 
(where $\lambda=M+\tilde{\omega}$), 
while the secular resonance variable 
$$\theta_{3}=\theta_{1}-\theta_{2}=
  \tilde{\omega}_{b}-\tilde{\omega}_{c}$$
librates about $180^{\circ}$ with a period of about 2800
years and an amplitude of 80 degrees (see Fig. 7). 
$\theta_{3}$ being only librating (i.e. without secular term),
the two orbital planes on average precess at the same rate.
This is a mechanism of secular resonance in periastron, which
as a consequence ensures the maintenance of the orbital topology
as well as the respective motions of the planets, that is to say
in the present case, the 2:1 mean motion resonance without
close approaches of the planets at their periapses.
In other words, this mechanism, when combined with
adequate relative positions of the planets on the initial
osculating orbits, can be understood as the condition for
the preservation of the dynamical stability related to the
2:1 mean motion resonance. Because the two variables 
$\theta_{1}$ and $\theta_{2}$ librate around $180^{\circ}$
and $0^{\circ}$, respectively (see Fig. 7), the lines of
apsides are anti-aligned. This situation is different from
the GJ$\thinspace$876 configuration in which the apsidal
lines are aligned with $\theta_{1}$, $\theta_{2}$, and
$\theta_{3}$ all librating about $0^{\circ}$ (Lee \& Peale
2002, Go\'{z}dziewski, Bois \& Maciejewski 2002). Let us
note that in the familiar Io-Europa 2:1 resonance, the very
small eccentricities lead to a geometry where $\theta_{1}$
is librating about $0^{\circ}$, while $\theta_{2}$ and
$\theta_{3}$ are librating about $180^{\circ}$ (Lee \& Peale
2002). In this case, conjunctions occur when Io is near
periapse and Europa near apoapse.

\placefigure{f7}

\clearpage

\begin{figure} 
\epsscale{0.8} 
%\plotone{f7.eps} 
\caption{Librations of the mean motion resonance variables 
$\theta_{1}$ and $\theta_{2}$, and of the secular resonance 
variable $\theta_{3}$. The dynamical behavior of these three
variables, derived from the integration of the geometrical
elements given for the HD$\thinspace$160691 planetary system 
(see the values in caption of Fig. 1), express a mechanism
of secular resonance in periastron with anti-aligned apsidal
lines.} 
\label{f7}
\end{figure} 

\clearpage

The whole stability mechanism for the HD$\thinspace$160691
system allows it to avoid close approaches between planets,
especially at their periapses (or near them). If it is not
the case, the dynamical behavior is wholly unstable since
the close approaches near periapses occur regularly
(see Fig. 3 where $M_{b}=M_{c}=0^{\circ}$).
The original orbital topology of the HD$\thinspace$160691
planetary system associated to the 2:1 mean motion
resonance can be written as $P_{c}(ap) - S - P_{b}(ap)$
\footnote{This 3-D topology with anti-aligned apsidal lines
is out of the scope of the various 2:1 resonance configurations
analyzed by Hadjidemetriou (2002) (2 types of configuration)
and Hadjidemetriou \& Psychogiou (in preparation) (4 types in
the planar case).},
which means that the planets {\it c} and {\it b} may be
considered as initially located at their apoastron around the
central star {\it S} (state 1). Taking into account the 2:1
mean motion resonance, the present topology is equivalent to 
$S - P_{c}(per) - P_{b}(ap)$~:
after one revolution of the planet {\it b}, the planet {\it c}
is at its periapse (state 2). This situation corresponds (by
taking into account the important amplitude of the secular
resonance variable $\theta_{3}$, namely $80\deg$) to the
nominal anomalies given in Table 1 
($M_{b}\sim 160\deg, M_{c}=0$). After one revolution more
of the planet {\it b}, the planets are again near their
apoastron (state 1). Besides, with the 2nd state, we
understand that the present topology may drive to a solid
stability where the sum of both eccentricities is
particularly important (i.e. $e_{b}+e_{c}>0.7$, 
high $e_{c}$ helps) (see Fig. 2).
Because of the high eccentricities of the orbits,
and despite relatively small semi-major axes, the relative
distances between the two planets may remain sufficiently
large over the whole evolutionary time scale of the system. 

\clearpage

\begin{figure}
\epsscale{1.0}
\plotone{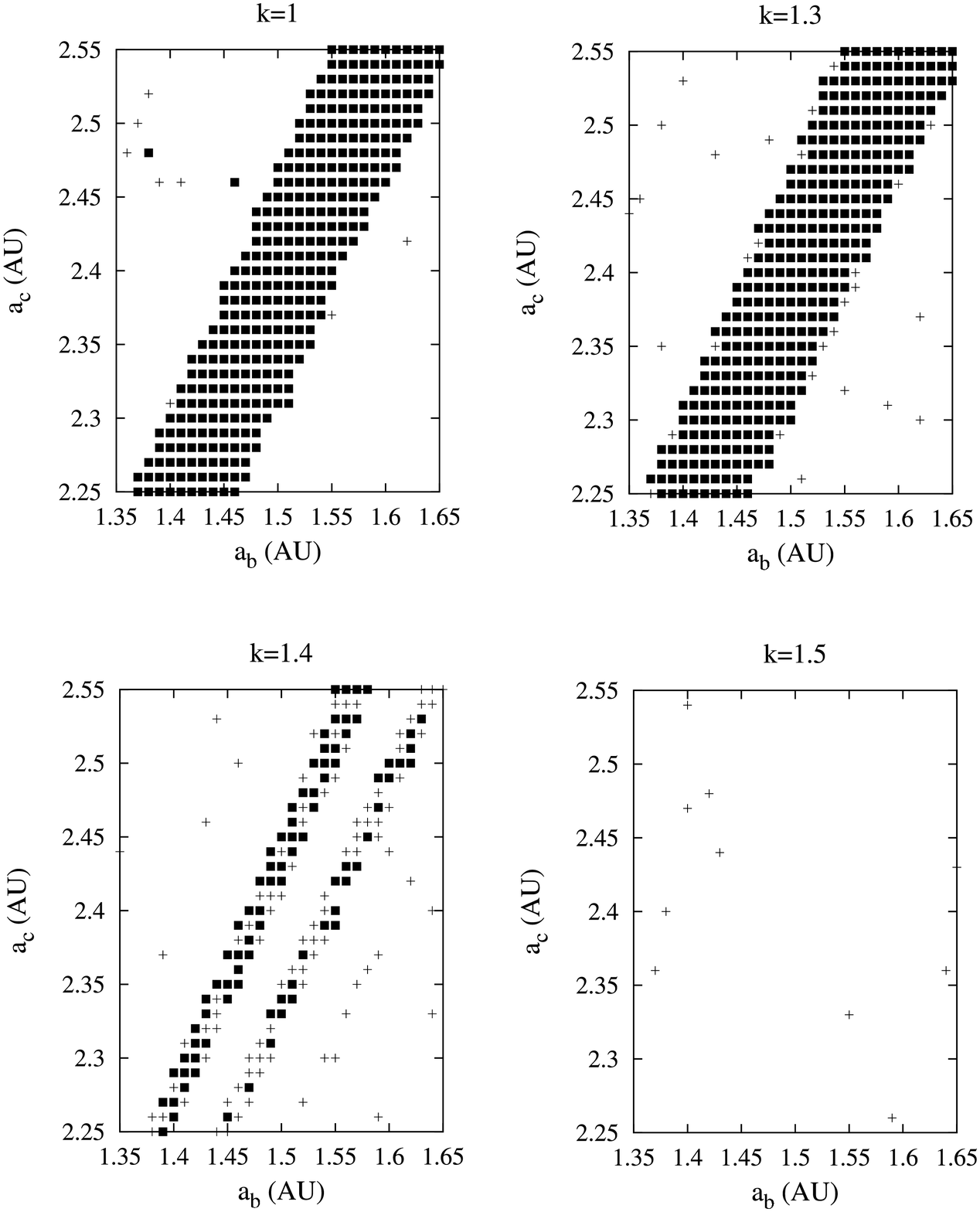} 
\caption{$[a_{b}, a_{c}]$ maps for different masses. $k$ is 
the multiple of the determined minimum mass of the planets~: 
(a) $k=1$, (b) $k=1.3$, (c) $k=1.4$, (d) $k=1.5$. 
The stable motion is represented by black squares 
(FLI $< 10^8$), "+" are plotted if $10^8 < $ FLI $< 10^{16}$, 
and the white area shows the chaotic regions (FLI $> 10^{16}$).} 
\label{f8}
\end{figure}

\placefigure{f8}

\clearpage
          
We have also tested the robustness of this mechanism when 
the values of the planetary masses were progressively 
increased. In Figure 8, computed by the FLI technique in 
the $[a_{b}, a_{c}]$ parameter space, the two masses are 
consecutively multiplied by a factor $k=$ 1 (a), 
1.3 (b), 1.4 (c), 1.5 (d). The total destruction of the 
stable structure is reached for $k=1.5$, which 
corresponds to a line-of-sight inclination factor 
$\sin i_{l}=2/3$ (i.e. $i_{l}=41.8^{\circ}$), as we know
only the lower limit of planetary masses $m_{0}$ from the
function $m_{0}=m_{P}\sin i_{l}$ where $m_{P}$ is the real
mass. This result is in agreement with a numerical analysis
by Kiseleva-Eggleton \& Bois (2001) of multi-planetary
systems of $\upsilon$ And-type which showed that the
dynamical regime, and consequently the lifetime of the
system, depends strongly on the mass hierarchy as well as
on the absolute values of planetary masses. 
In other words, assuming that the observed planetary system 
is dynamically stable, we could probably determine a window 
for the possible values of line-of-sight inclinations $i_{l}$, 
and as a consequence give upper limits on the planetary masses. 

\section{Conclusion} 

Using the MEGNO technique of global dynamics analysis we 
scanned the most relevant cases of the orbital parameter 
space for the HD$\thinspace$160691 planetary system. We have
found the existence of a stability zone ruled by a mechanism
which involves angular elements of the system. This stability
zone is indeed due to the 2:1 mean motion resonance coupled
with adequate relative positions of the planets on their orbits
avoiding close approaches at their periastron, the two apsidal 
lines being anti-aligned. The mechanism is not lost during the
dynamical evolution of the system due to an apsidal secular 
resonance~: the mean motion resonance variables are librating 
while the longitudes of periapse on average precess at the
same rate.
                   
We conclude that in order to be dynamically stable, 
the HD$\thinspace$160691 planetary system has to satisfy 
the following conditions~: (1) a 2:1 mean motion resonance, 
(2) combined with an apsidal secular resonance, 
(3) in a configuration $P_{c}(ap) - S - P_{b}(ap)$ 
(i.e. an apsidal antialignment), 
(4) and specific conditions on the respective sizes of the
eccentricities. High eccentricity for the outer orbit 
($e_{c} > 0.52$) is the most probable necessary condition,
while the eccentricity of the inner orbit $e_b$ becomes
relatively unimportant when $e_{c} > 0.7$.
These four conditions, taking also into account various
relative inclinations between the two orbits, determine
the dynamical behavior of the system in such a way
that the planets are never too close to each other.
In the end, the HD$\thinspace$160691 system, where the
resonance variables $\theta_{1}$ and $\theta_{3}$ librate
about $180^{\circ}$ while $\theta_{2}$ librates about
$0^{\circ}$, has revealed resources of the 2:1 orbital
resonances that have not been observed nor analyzed before.
The present orbital topology [$P_{c}(ap) - S - P_{b}(ap)$]
combined with the 2:1 orbital resonance is indeed more 
wide-ranging than the particular case of the 
HD$\thinspace$160691 planetary system. It is a new 
theoretical possibility suitable for a stable regime 
despite relatively small semi-major axes with respect
to the important masses in interactions.\footnote{Whereas
we are finishing the present paper, we learn that an apsidal
antialignment has been proposed for the dynamical stability
of the HD$\thinspace$82943 planetary system related to a 2:1 
orbital resonance (Ji \etal 2003).}

Combining our MEGNO maps (confirmed with the FLI method) 
for different pairs of parameters, it is possible to 
converge towards stability conditions related to different 
combinations of the angular parameters. As the observational 
determination of the elements of the HD$\thinspace$160691 
system is far from being finalized, we hope that the maps 
presented here will be useful for testing both future 
observations and different parameter fitting techniques.
In this respect, let us mention the fitting method of Laughlin
\& Chambers (2001), suitable for resonant interactions between
the planets and where the true masses can be determined by
eliminating the indeterminacy in $\sin i_{l}$ inherent in
fits that assume independent Keplerian motions. Let us
note here that we have already tested some of the new 
fits for HD$\thinspace$160691 obtained by Eugenio Rivera 
(private communication), and found that the only stable 
systems with $k < 2$ that do not meet our stability 
conditions would be ones with a very massive substellar 
distant companion ($m_c \sim 50 m_J$) on a very large 
orbit ($a_c \sim 30$ AU). This type of system is very 
different from the systems with Jupiter-mass planets 
on close orbits that we have discussed in this paper.   
           
\acknowledgments 

LK-E thanks the University Bordeaux 1 for a short-term 
research fellowship in Bordeaux and John Hadjidemetriou
for the text of his yet unpublished paper. EB and LK-E thank 
Rudolf Dvorak and the Institute of Astronomy, University of
Vienna, for hospitality. 
EP-L wishes to acknowledge the support 
by the Austrian FWF (Hertha Firnberg Project T122). 
The authors thank Eugenio Rivera for providing us with his 
latest unpublished orbital fits, and for useful discussions.
We thank Peter Eggleton for his comments and English corrections, 
and the referee Georges Contopoulos whose questions and comments 
greatly helped to improve the paper.

\end{document}